\documentclass[aps,prd,superscriptaddress]{revtex4}
\tighten

\newcommand{\ra}{\rightarrow}

\newcommand{\be}{\begin{equation}}
\newcommand{\ee}{\end{equation}}
\newcommand{\bea}{\begin{eqnarray}}
\newcommand{\eea}{\end{eqnarray}}

\newcommand{\s}{\smallskip}

\newcommand{\bc}{\begin{center}}
\newcommand{\ec}{\end{center}}
\newcommand{\bu}{\begin{underline}}
\newcommand{\eu}{\end{underline}}

\newcommand{\ty}{\textstyle}
\begin{document}
\draft
\title{Using gauge-invariant variables in QCD}\thanks{Contribution to Proceedings of the $8^{\rm th}$ International 
Wigner Symposium, May 27-30, 2003, New York, NY.}
\author{Kurt Haller\thanks{Kurt.Haller@uconn.edu}} 
\address{Department of Physics, University of Connecticut, Storrs, Connecticut
06269}
\begin{abstract}
The Weyl-gauge ($A_0^a=0)$ QCD Hamiltonian is unitarily transformed to a representation in 
which it is expressed entirely in terms of gauge-invariant quark and gluon fields. 
In a subspace of gauge-invariant states we have constructed that implement the non-Abelian Gauss's
law, this unitarily transformed Weyl-gauge Hamiltonian can be further transformed  and,
under appropriate circumstances, can be identified with the QCD Hamiltonian in the Coulomb gauge.
To circumvent the problem that this  Hamiltonian, which is expressed entirely in terms of gauge-invariant 
variables, must be used with nonnormalizable complicated states that implement the non-Abelian Gauss's law,
we demonstrate an isomorphism that materially facilitates the  application of this Hamiltonian to a variety of physical
processes, including the evaluation of $S$-matrix elements. This 
isomorphism relates the gauge-invariant representation
of the Hamiltonian and the required set of gauge-invariant states 
to a Hamiltonian of the same functional 
form but dependent on ordinary unconstrained Weyl-gauge fields operating within a space of `standard' 
perturbative states. The fact that the gauge-invariant 
chromoelectric field is not hermitian  has important implications
for the functional form of the Hamiltonian finally obtained. When this nonhermiticity is taken into account, 
the `extra' perturbative terms in Christ and Lee's Coulomb-gauge Hamiltonian are natural outgrowths of the formalism. When
this nonhermiticity is neglected, the Hamiltonian used in the earlier work of Gribov and others results. 
\end{abstract}
\maketitle

\narrowtext

Quantum Chromodynamics (QCD) exists in a variety of gauges; and the appearance of the Hamiltonian is profoundly different in 
the various gauges. In Quantum Electrodynamics (QED), in which a similar situation prevails, it has been possible to show 
that this dependence of the Hamiltonian on the gauge stems largely from the fact that the charged particles created from the vacuum by the 
charged fields and their conjugate momenta are not the same in different gauges --- for example, in 
Coulomb-gauge QED, the spinor fields create
electrons accompanied by their spherically symmetric gauge fields, while the same operators in the covariant gauges create
`bare' Dirac electrons unaccompanied by any electric or magnetic fields. In the spatial axial gauge, the electron is accompanied
by an electric field that suffices for satisfying Gauss's law, but that is severely anisotropic. It is 
this that causes the form of
Hamiltonians to depend so greatly on the choice of gauge. 
It has been possible to show that when differences in the 
definition of charged-particle creation and annihilation operators are taken into account by imposing gauge-invariance 
(or isotropy, in the case of the spatial axial gauge), the Hamiltonians for QED in different  gauges become remarkably similar.~\cite{khelqed} In this paper we will 
review work in which a similar result is obtained for QCD for the Weyl and Coulomb gauges. By 
imposing gauge invariance on the
operator-valued fields in Weyl-gauge QCD, we bring the Hamiltonian to a form that is 
almost identical to the Coulomb-gauge Hamiltonian, 
but that still describes the Weyl gauge. In the process, we also clarify some features of the Coulomb-gauge QCD Hamiltonian. 
\s

The basis for this work is the construction of states $|{\hat \Psi}\rangle$ 
that implement the non-Abelian Gauss's law, ${\hat {\cal G}}^a({\bf{r}})|{\hat \Psi}\rangle=0$, where 
the `Gauss's law operator'
${\hat {\cal G}}^a({\bf{r}})$ is related to its constituent fields in the
Weyl ($A^a_0=0$) gauge as shown by 
\be{\hat {\cal G}}^a({\bf{r}})=\overbrace{\partial_i\Pi^a_i({\bf{r}})+
\underbrace{gf^{abc}A^b_i({\bf{r}}){\Pi}^c_i({\bf{r}})}_{J^a_0({\bf{r}})}}^{D_i\Pi^a_i({\bf{r}})\,{\equiv}\,
{\cal G}^a({\bf{r}})}+j^a_0({\bf{r}}),
\ee
where ${\Pi}^a_i$ is the momentum conjugate to $A^a_i$, and 
$j^a_0({\bf{r}})=g\psi^{\dagger}({\bf{r}})\frac{{\lambda}^a}{2} \psi({\bf{r}}).$ We have 
constructed such a set of states for the `pure glue' Gauss's law operator 
${\cal G}^a({\bf{r}})=D_i\Pi^a_i({\bf{r}})$ by solving the equation
\be
\left\{\,\partial_{i}\Pi^a_{i}({\bf{r}}) + J_{0}^{a}({\bf{r}})\,\right\}
{\Psi}\,|{\phi}\rangle =0,
\ee
where we have chosen $|{\phi}\rangle$ to be a state that obeys the Abelianized Gauss's law, 
$\partial_i\Pi^a_i({\bf{r}})|{\phi}\rangle=0$.~\cite{CBH2} This requires that the operator-valued
$\Psi$ obey the `soft' equation, 
\be
[\,\partial_{i}\Pi^a_{i}({\bf{r}}),\,{\Psi}\,]\approx-J_{0}^{a}({\bf{r}}){\Psi},
\label{eq:soft}
\ee
which is only valid when both sides of Eq.~(\ref{eq:soft}) act on $|{\phi}\rangle$ states, so 
that any operators with $\partial_{i}\Pi^a_{i}$ on their extreme right-hand sides drop out.\s

Eq.~(\ref{eq:soft}) is suggestive of a solution for $\Psi$ in the form of an exponential. But a form 
$\Psi=\exp(\alpha)$ is not a viable candidate solution because, in the commutator of $\partial_{i}\Pi^a_{i}$ with a
representative term in $\exp(\alpha)$, 
$${\ty \frac{1}{n!}}\left[\partial_{i}\Pi^a_{i}({\bf{r}})\,,\alpha^n\right]={\ty \frac{1}{n!}}\sum_{i=0}^n\alpha^i
\left[\partial_{i}\Pi^a_{i}({\bf{r}})\,,\alpha\right]\alpha^{n-i-1}\,,
$$
$[\partial_{i}\Pi^a_{i}({\bf{r}})\,,\alpha]$ will not generally commute with $\alpha$, so that a simple
exponential function will not satisfy Eq.~(\ref{eq:soft}). In Ref.~\cite{CBH2}, we have found a solution in the
form 
\be
\Psi={\|}\,\exp({\cal{A}})\,{\|}\;,
\ee 
in which 
\be{\cal{A}}=
i{\int}d{\bf{r}}\;\overline{{\cal{A}}_{i}^{\gamma}}({\bf{r}})\;
\Pi_i^{\gamma}({\bf{r}})
\ee
and 
$\overline{{\cal{A}}_{i}^{\gamma}}({\bf{r}})$ is the `resolvent field'. 
The resolvent field is an operator-valued functional of the gauge field $A^a_i$, 
and is independent of the conjugate momentum 
$\Pi_i^{\gamma}$. The resolvent field is
central to this work, and largely determines
the properties of gauge-invariant fields and of Weyl-gauge QCD represented in terms of 
those gauge-invariant fields. The $||\;\;||$-ordered product is defined so that   
all functionals of $A^a_i$ are to the left of all functionals of $\Pi^b_j\,.$\s

The `soft' equation~(\ref{eq:soft}) has been shown to be equivalent to the equation
\be
{\int}d{\bf r}\overline{{\cal A}_{j}^{\gamma}}({\bf r})V_{j}^{\gamma}({\bf r})=
\sum_{n=1}^\infty
{\textstyle\frac{ig^n}{n!}}{\int}d{\bf r}\;
\!\left\{\,\psi^{\gamma}_{(n)j}({\bf{r}})\,+f^{\vec{\alpha}\beta\gamma}_{(n)}\,
{\cal{M}}_{(n)}^{\vec{\alpha}}({\bf{r}})\,
\overline{{\cal{B}}_{(n) j}^{\beta}}({\bf{r}})\,\right\}\;
\!\!V_{j}^{\gamma}({\bf r})\;,
\label{eq:hard}
\ee
in which $\overline{{\cal Y}^{\alpha}}({\bf r})=
{\textstyle \frac{\partial_{j}}{\partial^{2}}\overline{{\cal A}_{j}^{\alpha}}({\bf r})}$,~
${\cal{M}}_{n}^{\vec{\alpha}}({\bf{r}})
=\!\!\prod_{i=1}^n
\overline{{\cal Y}^{\alpha\,(i)}}({\bf{r}})$,~ and $\;\overline{{\cal B}_{(n) i}^{\beta}}({\bf r})=
a_i^{\beta}({\bf r})+\,
\left(\,\delta_{ij}-{\textstyle\frac{n}{(n+1)}}
{\textstyle\frac{\partial_{i}\partial_{j}}{\partial^{2}}}\,\right)
\overline{{\cal A}_{i}^{\beta}}({\bf r})$
where $\psi^{\gamma}_{(n)j}({\bf{r}})$ is a `source' term dependent on the gauge field and 
$a_i^{\beta}({\bf r})$ is the transverse part~of the gauge field);~\cite{CBH2}
$f^{\vec{\alpha}\beta\gamma}_{(n)}$
is the chain of structure constants 
$$f^{\vec{\alpha}\beta\gamma}_{(n)}=f^{\alpha[1]\beta b[1]}\,
\,f^{b[1]\alpha[2]b[2]}\,f^{b[2]\alpha[3]b[3]}\cdots f^{b[n-2]\alpha[n-1]b[n-1]}
f^{b[n-1]\alpha[n]\gamma}\;.$$
Eq.~(\ref{eq:hard}) can easily be seen to be a nonlinear integral equation for the resolvent field,
since $\overline{{\cal A}_{j}^{\gamma}}$ appears in $\overline{{\cal{B}}_{(n) j}^{\beta}}$ and in
highly multilinear forms in ${\cal{M}}_{(n)}^{\vec{\alpha}}$ on the right-hand side of this equation. 
Eq.~(\ref{eq:hard}) has a formal iterative solution;~\cite{CBH2,bellchenhall} when
Eq.~(\ref{eq:hard}) is specialized to the SU(2) case, and the resolvent field is given a $c$-number `hedgehog'
realization, noniterated solutions can be found that demonstrate topological sectors with half-integral 
as well as integral winding  numbers.~\cite{HCC} \s

The mathematical apparatus developed to construct $\Psi$ also is instrumental in the construction of 
gauge-invariant fields. We observe that the Gauss's law operator ${\hat {\cal G}}^a({\bf{r}})$ and its 
pure glue version, ${\cal G}^a({\bf{r}})=\partial_i\Pi^a_i({\bf{r}})+{gf^{abc}A^b_i({\bf{r}}){\Pi}^c_i({\bf{r}})}$,
are unitarily related as shown by
\be\hat{\cal{G}}^a({\bf{r}})={\cal{U}}_{\cal{C}}\,
{\cal{G}}^a({\bf{r}})\,{\cal{U}}^{-1}_{\cal{C}}\;\;\;\mbox{\rm where}\:\:\:
{\cal{U}}_{\cal{C}}=e^{{\cal C}_{0}}
e^{\bar {\cal C}}
\label{eq:unitary}
\ee

$$\mbox{\rm and where}\:\:\:{\cal C}_{0}=i\,\int d{\bf{r}}\,
{\textstyle {\cal X}^{\alpha}}({\bf r})\,j_{0}^{\alpha}({\bf r})
\;\;\mbox{and}\;\;\;
{\bar {\cal C}}=i\,\int d{\bf{r}}\,
\overline{{\cal Y}^{\alpha}}({\bf r})\,j_{0}^{\alpha}({\bf r})\;.$$
With this unitary equivalence, we are free to interpret ${\cal G}^a({\bf{r}})$ not as the pure glue
Gauss's law operator, but as the {\em complete} Gauss's law operator in a new representation, which we call the
${\cal N}$ representation. In this ${\cal N}$ representation, it is manifest that the spinor (quark) field 
$\psi$ is gauge-invariant since it trivially commutes with ${\cal G}^a$, the generator of infinitesimal 
gauge transformations. To find the gauge-invariant quark field in the original representation --- the
{\cal C} representation --- we simply invert the transformation in Eq.~(\ref{eq:unitary}) to obtain
\be
{\psi}_{\sf GI}({\bf{r}})=V_{\cal{C}}({\bf{r}})\,\psi ({\bf{r}})
\;\;\;\mbox{\small and}\;\;\;
{\psi}_{\sf GI}^\dagger({\bf{r}})=
\psi^\dagger({\bf{r}})\,V_{\cal{C}}^{-1}({\bf{r}}),
\label{eq:gtspin}
\ee
 $$\mbox{\rm where}\;\;\;V_{\cal{C}}({\bf{r}})=
\exp\left(\,-ig{\overline{{\cal{Y}}^\alpha}}({\bf{r}})
{\textstyle\frac{\lambda^\alpha}{2}}\,\right)\,
\exp\left(-ig{\cal X}^\alpha({\bf{r}})
{\textstyle\frac{\lambda^\alpha}{2}}\right)\;.$$
Under a gauge transformation that transforms $\psi$ as 
\be
{\psi}{\rightarrow}\exp(-i{\omega}^{\gamma}\,
\textstyle{\frac{{\lambda}^{\gamma}}{2}})\psi\,,
\ee
$V_{\cal C}$ transforms as 
\be
V_{\cal C}{\rightarrow}V_{\cal C}\exp(i{\omega}^{\gamma}\,
\textstyle{\frac{{\lambda}^{\gamma}}{2}})\,,
\label{eq:vc}
\ee
so that ${\psi}_{\sf GI}({\bf{r}})$ is strictly gauge-invariant.~\cite{mario} 
It is clear that $V_{\cal C}$ has the {\em   formal structure} of a gauge transformation
because there is always a ${\cal Z}^\alpha$ such that
\be
V_{\cal C}=\exp\left[-ig{\cal Z}^\alpha
{\textstyle\frac{{\lambda}^\alpha}{2}}\right] 
\;\;\;\;\mbox{and}\;\;\;\;
\exp\left[-ig{\cal Z}^\alpha
{\textstyle\frac{{\lambda}^\alpha}{2}}\right]=
\exp\left[-ig{\overline {\cal Y}^\alpha}
{\textstyle\frac{{\lambda}^\alpha}{2}}\right]\,
\exp\left[-ig{\cal X}^\alpha
{\textstyle\frac{{\lambda}^\alpha}{2}}\right]\,.
\ee
$V_{\cal C}$ is not, in fact, a gauge transformation because ${\cal Z}^\alpha$, rather than being a gauge function
by which a field is transformed, is itself an operator-valued functional of the gauge field that is subject to
gauge transformations as shown in Eq.~({\ref{eq:vc}}). Nevertheless, the formal structure of Eq.~({\ref{eq:gtspin}})
enables us to construct the gauge-invariant gauge (gluon) field 
\be
{\sf A}_{{\sf GI}\,i}({\bf{r}})=[\,A_{{\sf GI}\,i}^{b}({\bf{r}})\,{\textstyle\frac{\lambda^b}{2}}\,]
=V_{\cal{C}}({\bf{r}})\,[\,A_{i}^b({\bf{r}})\,
{\textstyle\frac{\lambda^b}{2}}\,]\,
V_{\cal{C}}^{-1}({\bf{r}})
+{\textstyle\frac{i}{g}}\,V_{\cal{C}}({\bf{r}})\,
\partial_{i}V_{\cal{C}}^{-1}({\bf{r}})\;,
\label{eq:AdressedAxz}
\ee
or, equivalently,
\be
A_{{\sf GI}\,i}^{b}({\bf{r}})=
A\,_{i}^{b\,{\sf T}} ({\bf{r}}) +
\left[\delta_{ij}-{\textstyle\frac{\partial_{i}\partial_j}
{\partial^2}}\right]\,\overline{{\cal A}_{i}^b} ({\bf{r}})
=
\left[\delta_{ij}-{\textstyle\frac{\partial_{i}\partial_j}
{\partial^2}}\right]\,\left\{A_b^i+\overline{{\cal A}_{i}^b}({\bf{r}})\right\}\,.
\ee
It is also possible to define a gauge-invariant negative chromoelectric field ${\Pi}^a_{{\sf GI}\,i}({\bf{r}})$
\be
{\Pi}^a_{{\sf GI}\,i}({\bf{r}})=R_{ab}({\bf{r}})\Pi^b_{i}({\bf{r}})\;\;\;\mbox{\rm where}\:\:\:
R_{ab}={\textstyle\frac{1}{2}}{\sf Tr}[{\lambda^a}V_{\cal{C}}{\lambda^b}V_{\cal{C}}^{-1}]\,.
\label{eq:gipi}
\ee
$R_{ab}$ and $\Pi^b_{i}$ do not commute, so that ${\Pi}^a_{{\sf GI}\,i}$ is not hermitian. 
For the SU(2) case, these gauge-invariant 
quantities have been shown to obey the commutation rules~\cite{HGrib} 
\be
\left[{\Pi}_{{\sf GI}\,i}^{\alpha}({\bf x})\,,
{\Pi}_{{\sf GI}\,j}^{\beta}({\bf y})\right]=
ig\left\{{\partial_i}{\cal D}^{{\alpha}h}({\bf x},{\bf y}){\epsilon}^{h\gamma\beta}
{\Pi}_{{\sf GI}\,j}^{\gamma}({\bf y})-{\partial_j}{\cal D}^{{\beta}h}({\bf y},{\bf x}){\epsilon}^{h\gamma\alpha}
{\Pi}_{{\sf GI}\,i}^{\gamma}({\bf x})\right\},
\label{eq:comma}
\ee
\be
\mbox{\rm and}\:\:\:\left[{\Pi}_{{\sf GI}\,j}^b({\bf y})\,,A^a_{{\sf GI}\,i}({\bf{x}})\right]=
-i\left\{\delta_{ab}\delta_{ij}{\delta}({\bf y}-{\bf x})-\partial_j{\cal D}^{bh}({\bf y},{\bf x})
\stackrel{\leftarrow}D_i^{\,ha}\!\!\!({\bf x})\right\}
\label{eq:commb}
\ee
which agree with those given by Schwinger for Yang-Mills theory in the Coulomb gauge
{\em modulo} operator order --- Schwinger's fields are symmetrized, whereas we 
never make {\em ad hoc} changes to operator order.~\cite{schwingera}\s

We can express the Weyl gauge Hamiltonian 
\be
H=\int d{\bf r} \left\{\ {\textstyle \frac{1}{2}
\Pi^{a}_{i}({\bf r})\Pi^{a}_{i}({\bf r})
+ \frac{1}{4}} F_{ij}^{a}({\bf r}) F_{ij}^{a}({\bf r})+
{\psi^\dagger}({\bf r})
\left[\,\beta m-i\alpha_{i}
\left(\,\partial_{i}-igA_{i}^{a}({\bf r})
{\textstyle\frac{\lambda^\alpha}{2}}\,\right)\,\right]
\psi({\bf r})\right\}\,
\label{eq:HQCDC}
\ee
in terms of gauge-invariant fields by transforming it from the ${\cal C}$ to the ${\cal N}$
representation. In this process we use the identities 
$$
R_{aq}R_{bq}=\delta_{ab},\;\;f^{duv}R_{ua}R_{vb}=f^{abq}R_{dq},\;\;\mbox{and}\;\;
\partial_iR_{ba}=-f^{uvb}R_{ua}P_{vi}\;\;\mbox{where}\;\;P_{vi}=-i{\sf Tr}[\lambda^vV_{\cal{C}}\,
\partial_{i}V_{\cal{C}}^{-1}]
$$
\be
{\rm and}\:\:\:{\cal U_C}^{-1}({\bf{x}})\Pi_i^a({\bf{x}})\,{\cal U_C}({\bf{x}})=
\Pi_i^a({\bf{x}})-R_{ba}({\bf{x}}){\partial^{({\bf{x}})}_i}{\int}d{\bf y}
{\cal D}^{\,bh}({\bf x},{\bf y})j_0^h({\bf y})\,,
\ee
which was demonstrated in Refs.~\cite{chen,BCH3}.    With Eq.~(\ref{eq:gipi}) and 
$$F_{{\sf GI}\,ij}^{a}({\bf r})=\partial_jA_{{\sf GI}\,i}^{a}({\bf r})-\partial_iA_{{\sf GI}\,j}^{a}({\bf r})-
g\epsilon^{abc}A_{{\sf GI}\,i}^{b}({\bf r})A_{{\sf GI}\,j}^{c}({\bf r})=R_{aq}({\bf r})F_{ij}^{q}({\bf r})$$
\be
 \mbox{\rm and with}\:\:\:\:\:\:J_{0\,({\sf GI})}^{a}({\bf r})=g{f}^{abc}A^b_{{\sf GI}\,i}({\bf{r}}) 
{\Pi}_{{\sf GI}\,i}^c({\bf r})\;\;\;\;\;\;\mbox{\rm and}
\label{eq:Jcolor}
\ee
\be
{\cal G}_{\sf GI}^{a}({\bf r})=\partial_i{\Pi}_{{\sf GI}\,i}^a({\bf r})+g{f}^{abc}A^b_{{\sf GI}\,i}({\bf r})
{\Pi}_{{\sf GI}\,i}^c({\bf r})=R_{ab}({\bf r}){\cal G}^{b}({\bf r})
\ee
we obtain
\bea
&&\!\!\!\!\!\!\!\!\!\!\!\!{\hat H}_{\sf GI}=\int d{\bf r}\left[ \ {\textstyle \frac{1}{2}}
\Pi^{a\,{\dagger}}_{{\sf GI}\,i}({\bf r})\Pi^{a}_{{\sf GI}\,i}({\bf r})
+  {\textstyle \frac{1}{4}} F_{{\sf GI}\,ij}^{a}({\bf r}) F_{{\sf GI}\,ij}^{a}({\bf r})-
{\psi^\dagger}({\bf r})\left(\beta m-i\alpha_{i}
\partial_{i}\right)\psi({\bf r})\right]+\nonumber \\
&&{\textstyle\frac{1}{2}}\int d{\bf x}d{\bf y}\left(J_{0\,({\sf GI})}^{a\,\dagger}({\bf x}) 
{\cal D}^{\,ab}({\bf x},{\bf y})j_0^b({\bf y})+j_0^b({\bf y})
\stackrel{\longleftarrow}{{\cal D}^{\,ba}}({\bf y},{\bf x})
J_{0\,({\sf GI})}^{a}({\bf x})\right)-\nonumber\\
&&{\textstyle\frac{1}{2}}{\int}d{\bf r}d{\bf x}d{\bf y}j_0^c({\bf y})
\stackrel{\longleftarrow}{{\cal D}^{\,ca}}({\bf y},{\bf r})
\partial^2{\cal D}^{\,ab}({\bf r},{\bf x})
j_0^b({\bf x})-{\int}d{\bf r}j^a_i({\bf r})A^a_{{\sf GI}\,i}({\bf{r}})+H_{\cal G}\,,
\label{eq:HQCDN}
\eea
where
\be
H_{\cal G}=-{\textstyle\frac{1}{2}}\int d{\bf x}d{\bf y}\big[{\cal G}_{\sf GI}^{a}({\bf x})
{\cal D}^{\,ab}({\bf x},{\bf y})j_0^b({\bf y})+
j_0^b({\bf y}){\cal D}^{\,ba}({\bf y},{\bf x}){\cal G}_{\sf GI}^{a}({\bf x})\big]\,.
\label{eq:HamGauss}
\ee
In Eqs.~(\ref{eq:comma}), (\ref{eq:commb}), (\ref{eq:HQCDN}) and (\ref{eq:HamGauss}), 
${\cal D}^{\,hb}({\bf y},{\bf x})$ is the inverse Faddeev-Popov operator defined by
\be
{\partial}{\cdot}D^{ah}_{({\bf y})}\,{\cal D}^{\,hb}({\bf y},{\bf x})=\delta_{ab}\delta({\bf y}-{\bf x})
\label{eq:FPd} 
\ee
where $D^{ah}={\partial_i}\delta_{ah}+gf^{a{\gamma}h}A^\gamma_{{\sf GI}\,i}$. Eq.~(\ref{eq:HQCDN}) has many features 
that suggest a Coulomb-gauge formulation --- the nonlocal interactions among color-charge densities and the 
fact that the interaction between the transverse $A^a_{{\sf GI}\,i}$ and the spatial current density 
$j^a_i({\bf r})$ is the only direct `contact' interactions between quarks and gluons. Eq.~(\ref{eq:HQCDN})
can be transformed further by noting that for states $|{\Psi}\rangle$ that implement Gauss's law in the ${\cal N}$
representation
\be
{\Pi}_{{\sf GI}\,i}^{c\,{\sf L}}|\Psi\rangle=-{\textstyle \frac{\partial_i}
{\partial^2}}J_{0\,({\sf GI})}^{c}|\Psi\rangle\;\;\;\;\;\;\mbox{\rm so that}
\ee   
\be
J_{0\,({\sf GI})}^{a}|\Psi\rangle=g{f}^{abc}A^b_{{\sf GI}\,i}\left({\Pi}_{{\sf GI}\,i}^{c\,{\sf T}}+
{\Pi}_{{\sf GI}\,i}^{c\,{\sf L}}\right)|\Psi\rangle=
\left\{J_{0\,({\sf GI})}^{a\,{\sf T}}-g{f}^{abc}A^b_{{\sf GI}\,i}
{\textstyle \frac{\partial_i}{\partial^2}}J_{0\,({\sf GI})}^{c}\right\}|\Psi\rangle\,,
\label{eq:jjt}
\ee
leading to the elimination of the longitudinal chromoelectric field as shown in 
\be
J_{0\,({\sf GI})}^b({\bf y})\approx\partial^2{\int}d{\bf x}{\cal D}^{\,ba}({\bf y},{\bf x})
J_{0\,({\sf GI})}^{a\,{\sf T}}({\bf x})\;\;\;\;\;\;\mbox{\rm and}\;\;\;\;\;\; 
J_{0\,({\sf GI})}^{b\,\dagger}({\bf y})\approx\partial^2{\int}d{\bf x}
J_{0\,({\sf GI})}^{a\,{\sf T}\,\dagger}({\bf x})
\stackrel{\longleftarrow}{{\cal D}^{\,ba}}({\bf y},{\bf x})\,,
\label{eq:jfpadj}
\ee
where $J_{0\,({\sf GI})}^{a\,{\sf T}}=g{f}^{abc}A^b_{{\sf GI}\,i}{\Pi}_{{\sf GI}\,i}^{c\,{\sf T}}$ and
$J_{0\,({\sf GI})}^{a\,{\sf T}\,\dagger}$ represents the hermitian 
adjoint of $J_{0\,({\sf GI})}^{a\,{\sf T}}$ and where, in this case, $\approx$ indicates that the equations only
hold within a space of $|\Psi\rangle$ states that implement the non-Abelian Gauss's law. 
Using these relations, we obtain a Hamiltonian that 
operates in the `physical' space of states that implement Gauss's law --- the $|\Psi\rangle$ states --- and
discard $H_{\cal G}$, which annihilates these states. The resulting `physical' Hamiltonian is 
\bea
&&\!\!\!\!\!\!\!\!\!\!({\hat H}_{\sf GI})_{\mbox{phys}}=\int\!d{\bf r}\left[ \ {\textstyle \frac{1}{2}}
\Pi^{a\,{\sf T}\,\dagger}_{{\sf GI}\,i}({\bf r})\Pi^{a\,{\sf T}}_{{\sf GI}\,i}({\bf r})
+  {\textstyle \frac{1}{4}} F_{{\sf GI}\,ij}^a({\bf r}) F_{{\sf GI}\,ij}^{a}({\bf r})+
{\psi^\dagger}({\bf r})\left(\beta m-i\alpha_{i}
\partial_{i}\right)\psi({\bf r})\right]-{\int}d{\bf r}j^a_i({\bf r})A^a_{{\sf GI}\,i}({\bf{r}})\nonumber\\
&&\;\;\;\;\;\;\;\;\;\;\;\;-{\ty \frac{1}{2}}\int\!d{\bf r}d{\bf x}d{\bf y}\left(j_0^b({\bf x}))+
J_{0\,({\sf GI})}^{b\,{\sf T}\,\dagger}({\bf x})\right)
\stackrel{\longleftarrow}{{\cal D}^{\,ab}}({\bf r},{\bf x}){\partial^2}
{\cal D}^{\,ac}({\bf r},{\bf y})\left(j_0^c({\bf y}))+
J_{0\,({\sf GI})}^{c\,{\sf T}}({\bf y})\right)\,.
\label{eq:Heff}
\eea
$({\hat H}_{\sf GI})_{\mbox{phys}}$ is very nearly identical to Gribov's Hamiltonian except that 
Gribov seems to have implicitly assumed that the transverse chromoelectric field is hermitian.~\cite{gribov} But
in fact, it is not, as can be seen from 
\be
\Pi_{{\sf GI}\,j}^{b\,{\sf T}\,\dagger}({\bf y})-\Pi_{{\sf GI}\,j}^{b\,{\sf T}}({\bf y})=
ig^2f^{hcb}f^{{\delta}cs}\left(\delta_{j,\ell}-\frac{\partial^{({\bf y})}_j
\partial^{({\bf y})}_\ell}{\partial^2}\right)
{\int}{d{\bf z}}{ \frac{\partial}{{\partial}y_\ell}}\left(\frac{1}{4{\pi}|{\bf y}-{\bf z}|}\right)
A_{{\sf GI}\,k}^{{\delta}}({\bf z})\frac{\partial}{{\partial}z_k}{\cal D}^{\,sh}({\bf z},{\bf y})\,. 
\label{eq:pipiTherm}
\ee
In order to make effective use of $({\hat H}_{\sf GI})_{\mbox{phys}}$, it is necessary to remove the 
restriction that the very complicated and nonnormalizable $|\Psi\rangle$ states must be used. This has been 
done by demonstrating that~\cite{khren}
\be
A_{{\sf GI}\,j}^a({\bf r})\Psi\,|\phi_i\rangle=\Psi\,A_{j}^{a\,{\sf T}}({\bf r})|\phi_i\rangle
\;\;\;\;\;\;\mbox{\rm and}\;\;\;\;\;\;
\Pi_{{\sf GI}\,j}^{c\,{\sf T}}({\bf r})\Psi\,|\phi_i\rangle=
\Psi\,\Pi_{j}^{c\,{\sf T}}({\bf r})|\phi_i\rangle\,.
\label{eq:rb}
\ee
We could use Eq.~(\ref{eq:rb}) to eliminate the restriction to $|\Psi\rangle$ states, but the fact that there is 
no similar relation for $\Pi_{{\sf GI}\,j}^{b\,{\sf T}\,\dagger}$, the hermitian adjoint of 
$\Pi_{{\sf GI}\,j}^{b\,{\sf T}}$, is an impediment to that program. We address that problem by making use of the fact 
(proven in Appendix B of Ref.~\cite{khren}) that $\Pi_{{\sf GI}\,j}^{b\,{\sf T}\,\dagger}={\cal J}^{-1}
\Pi_{{\sf GI}\,j}^{b\,{\sf T}}{\cal J}$, where ${\cal J}$ is the Faddeev-Popov determinant. Since ${\cal J}$ is
hermitian, we can define ${\cal P}^{b\,{\sf T}}_{j}$, a `hermitianized' momentum conjugate to ${A}^a_{{\sf GI}\,i}$,
by 
\be
\Pi_{{\sf GI}\,j}^{b\,{\sf T}}({\bf r})={\cal J}^{\frac{1}{2}}{\cal P}^{b\,{\sf T}}_{j}({\bf r})
{\cal J}^{-{\frac{1}{2}}}\;\;\mbox{and}\;\;\Pi_{{\sf GI}\,j}^{b\,{\sf T}\,\dagger}({\bf r})=
{\cal J}^{-\frac{1}{2}}{\cal P}^{b\,{\sf T}}_{j}({\bf r})
{\cal J}^{{\frac{1}{2}}}\,.
\label{eq:hermpb}
\ee
By using Eq.~(\ref{eq:hermpb}), we are now able to express the Hamiltonian $({\hat H}_{\sf GI})_{\mbox{phys}}$ in a form in which 
it is expressed in terms of ${A}^a_{{\sf GI}\,i}$ and ${\cal P}^{b\,{\sf T}}_{j}$. We find that 
$({\hat { H}}_{\sf GI})_{\mbox{phys}}=[{\sf H}]_0+[{\sf H}]_1+[{\sf H}]_2$, where
\be
[{\sf H}]_0=\int\!d{\bf r}\left[ \ {\textstyle \frac{1}{2}}
{\cal P}^{b\,{\sf T}}_{j}({\bf r}){\cal P}^{b\,{\sf T}}_{j}({\bf r})
+  {\textstyle \frac{1}{4}} {\hat F}_{{\sf GI}\,ij}^a({\bf r}) {\hat F}_{{\sf GI}\,ij}^{a}({\bf r})+
{\psi^\dagger}({\bf r})\left(\beta m-i\alpha_{i}
\partial_{i}\right)\psi({\bf r})\right]\,,
\label{eq:h0h}
\ee
\bea
&&\!\!\!\!\!\!\!\!\![{\sf H}]_1=\int\!d{\bf r}\left\{gf^{abc}\partial_iA^a_{{\sf GI}\,j}({\bf r})
A^b_{{\sf GI}\,i}({\bf r})A^c_{{\sf GI}\,j}({\bf r})+ 
{\textstyle \frac{1}{4}}g^2f^{abc}f^{ab^{\prime}c^{\prime}}
A^b_{{\sf GI}\,i}({\bf r})A^c_{{\sf GI}\,j}({\bf r})A^{b^{\prime}}_{{\sf GI}\,i}({\bf r})
A^{c^{\prime}}_{{\sf GI}\,j}({\bf r})\right.\nonumber \\
&&\!\!\!\!\!\!\!\!\!-\left.j^a_i({\bf r})A^a_{{\sf GI}\,i}({\bf{r}})\right\}-
{\ty \frac{1}{2}}\int\!d{\bf r}d{\bf x}d{\bf y}\left(j_0^b({\bf x})+
{\bar {\sf J}}_{0\,({\sf GI})}^{b\,{\sf T}}({\bf x})\right)
\stackrel{\longleftarrow}{{\cal D}^{\,ba}}({\bf x},{\bf r}){\partial^2}
{\cal D}^{\,ac}({\bf r},{\bf y})\left(j_0^c({\bf y})+
{\bar {\sf J}}_{0\,({\sf GI})}^{c\,{\sf T}}({\bf y})\right)\,
\label{eq:h1h}
\eea
\nopagebreak
\bea
\mbox{and }\;\;\;\;&&[{\sf H}]_2={\cal U}+{\cal V}+{\ty \frac{1}{2}}\int\!d{\bf r}d{\bf x}d{\bf y}\left\{
i{\sf k}^b_0({\bf x})\stackrel{\longleftarrow}{{\cal D}^{\,ba}}({\bf x},{\bf r}){\partial^2}
{\cal D}^{\,ac}({\bf r},{\bf y})\left(j_0^c({\bf y})+
{\bar {\sf J}}_{0\,({\sf GI})}^{c\,{\sf T}}({\bf y}))\right)-\right.\nonumber\\
&&\left.\left(j_0^b({\bf x})+
{\bar {\sf J}}_{0\,({\sf GI})}^{b\,{\sf T}}({\bf x})\right)
\stackrel{\longleftarrow}{{\cal D}^{\,ba}}({\bf x},{\bf r}){\partial^2}
{\cal D}^{\,ac}({\bf r},{\bf y})i{\sf k}^c_0({\bf y})+
{\sf k}^b_0({\bf x})\stackrel{\longleftarrow}{{\cal D}^{\,ba}}({\bf x},{\bf r}){\partial^2}
{\cal D}^{\,ac}({\bf r},{\bf y}){\sf k}^c_0({\bf y})\right\}
\label{eq:h2h}
\eea
where ${\bar {\sf J}}_{0\,({\sf GI})}^{a\,{\sf T}}=gf^{abc}A^b_{{\sf GI}\,i}{\cal P}^{c\,{\sf T}}_{i}$ and is hermitian.
${\cal U}$ and ${\cal V}$ are complicated expressions that are given explicitly in Appendix E of Ref.~\cite{khren}.
The sum $[{\sf H}]_0+[{\sf H}]_1$ is identical in form to the Coulomb-gauge QCD Hamiltonian used
by Gribov,~\cite{gribov} as well as by numerous other authors who have followed him in using
this Hamiltonian. $[{\sf H}]_2$ consists of additional terms that are required because 
the transverse gauge-invariant negative chromoelectric field
$\Pi_{{\sf GI}\,j}^{p\,{\sf T}}$ is not hermitian. In Ref.~\cite{khren}, we prove that $[{\sf H}]_2$ accounts for 
the `extra' nonlocal interactions noted by Christ and Lee,~\cite{christlee} 
beyond those included in the Coulomb-gauge Hamiltonian 
used in earlier work, for example in Ref.~\cite{gribov}.
 We thus have been able to 
connect these extra nonlocal interactions with the nonhermiticity of the 
transverse ${\Pi}_{{\sf GI}\,j}^{b\,{\sf T}}$.\s

We are able to eliminate the restriction that this Hamiltonian only be used with the very complicated $|\Psi\rangle$ states that 
implement the non-Abelian Gauss's law, by demonstrating an isomorphism based on the fact that 
${\cal P}^{b\,{\sf T}}_{j}({\bf y})$ and $A_{{\sf GI}\,i}^{a}({\bf{x}})$ obey the same commutation rules
\be
\left[A_{{\sf GI}\,j}^{b}({\bf{y}})\,,A_{{\sf GI}\,i}^{a}({\bf{x}})\right]
=\left[{\cal P}^{b\,{\sf T}}_{j}({\bf y})\,,{\cal P}^{a\,{\sf T}}_{i}({\bf x})\right]=0,\;\;\;\;\mbox{and}\;\;\;\;
\left[{\cal P}^{b\,{\sf T}}_{j}({\bf y})\,,A_{{\sf GI}\,i}^{a}({\bf{x}})\right]
=-i\delta_{ab}\left(\delta_{ij}-\frac{\partial_i\partial_j}{\partial^2}\right)
\delta({\bf x}-{\bf y})
\ee
as do the transverse Weyl-gauge fields $\Pi_{j}^{b\,{\sf T}}({\bf y})$ and $A_{i}^{a\,{\sf T}}({\bf x})$,
We are therefore able to represent $A_{{\sf GI}\,i}^{a}({\bf{x}})$ and ${\cal P}^{b\,{\sf T}}_{j}({\bf y})$ in terms
of particle creation and annihilation operators in a completely parallel way as is standard for $A_{i}^{a\,{\sf T}}({\bf x})$ and  
$\Pi_{j}^{b\,{\sf T}}({\bf y})$, namely
\be
A_{{\sf GI}\,i}^c({\bf r})=\sum_{{\bf k},\,n}\frac{\epsilon_i{}^n({\bf k})}
{\sqrt{2k}}\left[{\alpha}^c_n({\bf k})e^{i{\bf
k\cdot r}} +{\alpha}_n^{c\,\dagger}({\bf k})e^{-i{\bf k\cdot r}}\right]
\;\;\;\mbox{and}\;\;\;
{\cal P}_{i}^{c\,{\sf T}}({\bf r})=-i\sum_{{\bf k},\,n}\epsilon_i{}^n({\bf
k})\sqrt{\frac{k}{2}}\left[{\alpha}^c_n({\bf k})e^{i{\bf
k\cdot r}} -{\alpha}_n^{c\,\dagger}({\bf k})e^{-i{\bf k\cdot r}}\right]
\label{eq:newrules}\ee
where $n$ is summed over two transverse helicity modes and
\be
\left[{\alpha}^a_n({\bf k})\,,{\alpha}_\ell^{b\,\dagger}({\bf q})\right]=
\delta_{n,\ell}\,\delta_{a,b}\,\delta_{{\bf k},{\bf q}}\;\;\mbox{and}
\;\;\left[{\alpha}^a_n({\bf k})\,,{\alpha}_\ell^{b}({\bf q})\right]=
\left[{\alpha}^{a\,\dagger}_n({\bf k})\,,{\alpha}_\ell^{b\,\dagger}({\bf q})\right]=0\,.
\label{eq:comalph}
\ee
Since an identical set of relations exists between the transverse Weyl-gauge fields $A_{j}^{c\,{\sf T}}({\bf r})$ 
and $\Pi_{j}^{c\,{\sf T}}({\bf r})$ and a set of excitation operators ${a}^a_n({\bf k})$ and 
${a}_n^{c\,\dagger}({\bf k})$ which also obey the commutation rules in Eq.~(\ref{eq:comalph}),
\be
A_{i}^{c\,{\sf T}}({\bf r})=\sum_{{\bf k},\,n}\frac{\epsilon_i{}^n({\bf k})}
{\sqrt{2k}}\left[{a}^c_n({\bf k})e^{i{\bf
k\cdot r}} +{a}_n^{c\,\dagger}({\bf k})e^{-i{\bf k\cdot r}}\right]
\;\;\;\mbox{and}\;\;\;
{\Pi}_{i}^{c\,{\sf T}}({\bf r})=-i\sum_{{\bf k},\,n}\epsilon_i{}^n({\bf
k})\sqrt{\frac{k}{2}}\left[{a}^c_n({\bf k})e^{i{\bf
k\cdot r}} -{a}_n^{c\,\dagger}({\bf k})e^{-i{\bf k\cdot r}}\right]\,,
\label{eq:oldrules}
\ee
we can use Eq.~(\ref{eq:rb}) to show that 
\be
{\alpha}_n^{c\,\dagger}({\bf k}){\cal J}^{-\frac{1}{2}}\Psi|\phi_i\rangle=
{\cal J}^{-\frac{1}{2}}\Psi\,{a}_n^{c\,\dagger}({\bf k})|\phi_i\rangle\;\;\;\mbox{\rm and}\;\;\;
{\alpha}_n^{c}({\bf k}){\cal J}^{-\frac{1}{2}}\Psi|\phi_i\rangle=
{\cal J}^{-\frac{1}{2}}\Psi\,{a}_n^{c}({\bf k})|\phi_i\rangle\,.
\label{eq:A2}
\ee
Any ${\alpha}_n^{c}({\bf k})$ will annihilate the gauge-invariant vacuum state 
${\cal J}^{-\frac{1}{2}}\Psi\,\Xi|0\rangle$, because the transverse excitation operators 
${a}_n^{c}({\bf k})$ and ${a}_n^{c\,\dagger}({\bf k})$ commute with $\Xi$ and  
${a}_n^{c}({\bf k})$ trivially annihilates $\Xi|0\rangle$.
\s

At this point, we can establish an isomorphism between two Fock spaces: The `standard' Weyl-gauge Fock space 
consists of 
\be
|{\bf k}\rangle={a}_n^{c\,\dagger}({\bf k})|0\rangle,\;\;\;\cdots\;\;\;|{\bf k}_1
\cdots{\bf k}_i\cdots{\bf k}_N\rangle={\ty \frac{1}{\sqrt{N!}}}\left[{a}_{n_{1}}^{{c_{1}}\,\dagger}({\bf k}_1)\cdots
{a}_{n_{i}}^{{c_{i}}\,\dagger}({\bf k}_i)\cdots{a}_{n_{N}}^{{c_{N}}\,\dagger}({\bf k}_N)\right]|0\rangle\,;
\label{eq:kn}
\ee 
and the gauge-invariant states that implement the non-Abelian Gauss's law can be represented as
\be
|{\bf {\bar k}}\rangle={\ty \frac{1}{C}}{\alpha}_n^{c\,\dagger}({\bf k})
{\cal J}^{-\frac{1}{2}}\Psi\Xi|0\rangle\;\;\;\cdots\;\;\;
|{\bf {\bar k}}_1\cdots{\bf {\bar k}}_i\cdots{\bf {\bar k}}_N\rangle={\ty \frac{1}{C\,\sqrt{N!}}}
\left[{\alpha}_{n_{1}}^{{c_{1}}\,\dagger}({\bf k}_1)\cdots
{\alpha}_{n_{i}}^{{c_{i}}\,\dagger}({\bf k}_i)\cdots{\alpha}_{n_{N}}^{{c_{N}}\,\dagger}({\bf k}_N)\right]
{\cal J}^{-\frac{1}{2}}\Psi\Xi|0\rangle
\ee
where $|0\rangle$ designates the perturbative vacuum annihilated by ${a}^c_n({\bf k})$ as well as by the
annihilation operators for quarks and antiquarks,  
$q_{{\bf p},s}$ and ${\bar q}_{{\bf p},s}$ respectively. The normalization constant
$C^{-1}$ must be introduced to compensate for the fact 
that $|C|^2=|{\cal J}^{-\frac{1}{2}}\Psi\Xi|0\rangle|^2=
\langle0|\Xi^\star\Psi^\star{\cal J}^{-1}\Psi\Xi|0\rangle$, which formally is a universal positive
constant, is not finite; and the state ${\cal J}^{-\frac{1}{2}}\Psi\Xi|0\rangle$  is not normalizable. 
However, once $C$ is introduced, the 
$|{\bf {\bar k}}_1\cdots{\bf {\bar k}}_i\cdots{\bf {\bar k}}_N\rangle$ 
states form a satisfactory Fock space that is gauge-invariant as well as 
isomorphic to the space of $|{\bf k}_1\cdots{\bf k}_i\cdots{\bf k}_N\rangle$ states.
We can now use 
Eq.~(\ref{eq:newrules}) to express $[{\sf H}]_0$ as
\be
[{\sf H}]_0=\sum_{{\bf k}, c}k{\alpha}_n^{c\,\dagger}({\bf k}){\alpha}_n^{c}({\bf k})+
\sum_{{\bf p},s}{\cal E}_{\bf p}\left(q_{{\bf p},s}^{\dagger}q_{{\bf p},s}+
{\bar q}_{{\bf p},s}^\dagger{\bar q}_{{\bf p},s}\right)\,.
\label{eq:h0spect}
\ee
In this form, $[{\sf H}]_0$ can be seen to describe the energy of non-interacting gauge-invariant
transverse gluons of energy $k$ and quarks and anti-quarks respectively of energy ${\cal E}_{\bf p}=\sqrt{m^2+|{\bf p}|^2}$.
We can also define another Hamiltonian, 
${\cal H}={\cal H}_0+{\cal H}_1+{\cal H}_2$, in which each component part is identical in form to 
$[{\sf H}]_0+[{\sf H}]_1+[{\sf H}]_2$ respectively, but with the substitutions 
$$
{\cal P}^{b\,{\sf T}}_{j}\ra\Pi^{b\,{\sf T}}_{j}\;\;\;\mbox{and}\;\;\;
A^a_{{\sf GI}\,i}{\ra} A^{a\,{\sf T}}_{i}
$$
everywhere --- including the replacement of $A^a_{{\sf GI}\,i}$ by $A^{a\,{\sf T}}_{i}$ 
in the inverse 
Faddeev-Popov operator ${\cal D}^{\,ab}({\bf x},{\bf y})$ ---
so that ${\cal H}$ is characteristic of the
Coulomb gauge, but nevertheless is a  functional of transverse Weyl-gauge unconstrained 
fields. For example, ${\cal H}_0$
is
\be
{\cal H}_0=\int d{\bf r} \left[\ {\textstyle \frac{1}{2}
\Pi^{a\,{\sf T}}_{i}({\bf r})\Pi^{a\,{\sf T}}_{i}({\bf r})
+ \frac{1}{4}} {\hat F}_{ij}^{a}({\bf r}) {\hat F}_{ij}^{a}({\bf r})+
{\psi^\dagger}({\bf r})
\left(\,\beta m-i\alpha_{i}
\partial_{i}\right)
\psi({\bf r})\right]\,
\label{eq:HQCDC}
\ee
where ${\hat F}_{ij}^a=\partial_jA^a_{i}-\partial_iA^a_{j}$. Using Eq.~(\ref{eq:oldrules}), we can express 
${\cal H}_0$ in the form
\be
{\cal H}_0=\sum_{{\bf k}, c}k{a}_n^{c\,\dagger}({\bf k}){a}_n^{c}({\bf k})+
\sum_{{\bf p},s}{\cal E}_{\bf p}\left(q_{{\bf p},s}^{\dagger}q_{{\bf p},s}+
{\bar q}_{{\bf p},s}^\dagger{\bar q}_{{\bf p},s}\right)\,,
\label{eq:h0wspec}
\ee 
 and use Eq.~(\ref{eq:A2}) to establish that 
\be
\!\!\!\![{\sf H}]_0
{\cal J}^{-\frac{1}{2}}\Psi\,\Xi|n\rangle=
{\cal J}^{-\frac{1}{2}}\Psi\,\Xi\,{\cal H}_0|n\rangle
\,,\;\;
[{\sf H}]_1
{\cal J}^{-\frac{1}{2}}\Psi\,\Xi|n\rangle=
{\cal J}^{-\frac{1}{2}}\Psi\,\Xi\,{\cal H}_1|n\rangle\,,
\;\;\;\mbox{\rm and}\;\;\;
[{\sf H}]_2
{\cal J}^{-\frac{1}{2}}\Psi\,\Xi|n\rangle=
{\cal J}^{-\frac{1}{2}}\Psi\,\Xi\,{\cal H}_2|n\rangle\,.
\label{eq:hotran2}
\ee
\s

We can use the relations between Weyl-gauge and gauge-invariant states we have established in the preceding discussion
to extend the isomorphism we have demonstrated to include scattering transition amplitudes. 
For this purpose, we define 
\be
 {\cal H}_{\rm int}={\cal H}_1+{\cal H}_2\;\;\;\mbox{and}\;\;\;[{\sf H}]_{\rm int}=[{\sf
H}]_1+[{\sf H}]_2\,.
\label{eq:hinttran}
\ee

The transition
amplitude between gauge-invariant states is given by 
\be
\overline{{\sf T}}_{f,i}=\frac{1}{C^2}{\langle}f|\Xi^\star\Psi^\star{\cal J}^{-\frac{1}{2}}
\left\{[{\sf H}]_{\rm int}+[{\sf H}]_{\rm int}
\frac{1}{\left(E_i-({\hat { H}}_{\sf GI})_{\mbox{phys}}+i\epsilon\right)}
[{\sf H}]_{\rm int}\right\}{\cal J}^{-\frac{1}{2}}\Psi\Xi|i\rangle\,,
\label{eq:hyb1}
\ee
where $|i\rangle$ and $|f\rangle$ each designate one of the $|n\rangle$ states; $|i\rangle$ represents an incident
and $|f\rangle$ a final state in a scattering process. With the results of the preceding discussion, we can express this as
\bea
\overline{{\sf T}}_{f,i}\!\!\!\!\!\!&&=\frac{1}{C^2}{\langle}f|\Xi^\star
\Psi^\star{\cal J}^{-1}\Psi\Xi|n\rangle{\langle}n|
\left\{ {\cal H}_{\rm int}+ {\cal H}_{\rm int}
\frac{1}{\left(E_i-{\cal H}_0- {\cal H}_{\rm int}+i\epsilon\right)}
{\cal H}_{\rm int}\right\}|i\rangle\nonumber \\
&&=\frac{1}{C^2}{\langle}0|\Xi^\star\Psi^\star{\cal J}^{-1}\Psi\Xi|0\rangle{\langle}f|
\left\{ {\cal H}_{\rm int}+ {\cal H}_{\rm int}
\frac{1}{\left(E_i-{\cal H}_0- {\cal H}_{\rm int}+i\epsilon\right)}
{\cal H}_{\rm int}\right\}|i\rangle
\label{eq:hyb1}
\eea
where we sum over the complete set of perturbative states $|n\rangle{\langle}n|$. The second line of
Eq.~(\ref{eq:hyb1}) follows from 
\bea
&&\frac{1}{C^2}{\langle}f|\Xi^\star
\Psi^\star{\cal J}^{-1}\Psi\Xi|n\rangle=\frac{1}{C^2}{\langle}0|a_{f}({\bf k}_f)\Xi^\star
\Psi^\star{\cal J}^{-1}\Psi{\Xi}a^\dagger_n({\bf k}_n)|0\rangle=\frac{1}{C^2}{\langle}0|\Xi^\star
\Psi^\star{\cal J}^{-\frac{1}{2}}\alpha_{f}({\bf k}_f)\alpha^\dagger_n({\bf k}_n)
{\cal J}^{-\frac{1}{2}}\Psi{\Xi}|0\rangle=\nonumber\\
&&
\delta_{f,n}{\delta}({\bf k}_f-{\bf k}_n)\frac{1}{C^2}{\langle}0|\Xi^\star
\Psi^\star{\cal J}^{-1}\Psi\Xi|0\rangle-\frac{1}{C^2}{\langle}0|\Xi^\star
\Psi^\star{\cal J}^{-\frac{1}{2}}\alpha^\dagger_n({\bf k}_n)\alpha_{f}({\bf k}_f)
{\cal J}^{-\frac{1}{2}}\Psi{\Xi}|0\rangle
\label{eq:hyb3}
\eea
and the observation that the last term on the second line of Eq.~(\ref{eq:hyb3}) vanishes trivially.
With the isomorphism of the states 
$|{\bf k}_1\cdots{\bf k}_i\cdots{\bf k}_N\rangle$ and 
$|{\bf {\bar k_1}}\cdots{\bf {\bar k_i}}\cdots{\bf {\bar k_N}}\rangle$ that we have established, 
\be
\overline{{\sf T}}_{f,i}=T_{f,i}
\label{eq:tt}
\ee
where 
\be
T_{f,i}={\langle}f|
\left\{ {\cal H}_{\rm int}+ {\cal H}_{\rm int}
\frac{1}{\left(E_i-{\cal H}_0- {\cal H}_{\rm int}+i\epsilon\right)}
 {\cal H}_{\rm int}\right\}|i\rangle
\label{eq:Treg}
\ee
is a transition amplitude 
that can be evaluated with Feynman graphs and rules, because it is based on `standard' 
perturbative states that are not required to implement Gauss's law and need not be gauge-invariant.\s

The formulation of Weyl-gauge QCD in terms of gauge-invariant fields leads to the Hamiltonian $({\hat { H}}_{\sf GI})_{\mbox{phys}}$
given in Eqs.~(\ref{eq:h0h})-(\ref{eq:h2h}) and the demonstration of physical equivalence to Coulomb-gauge QCD.
But the demonstration that this Hamiltonian is useful in actual calculations depends on the further demonstration of the
isomorphism that identifies $({\hat { H}}_{\sf GI})_{\mbox{phys}}$ with ${\cal H}={\cal H}_{0}+{\cal H}_1+{\cal H}_2$
and the development leading up to Eq.~(\ref{eq:tt}).\s

This research was supported by the Department of Energy under Grant No.~DE-FG02-92ER40716.00.

\end{document}